\documentclass[5p,twocolumn,final]{elsarticle}
\biboptions{numbers,sort&compress}
\usepackage[version=3]{mhchem}  % \ce{...} for chem equations
\usepackage{amssymb}

\newcommand{\textcite}[1]{\citet{#1}}
\usepackage[warning,np]{numprint}
\npdecimalsign{\ensuremath{.}}
\npthousandsep{\,}
\npproductsign{\times}
\npfourdigitnosep
\usepackage{booktabs} % better tables
\graphicspath{{.}{./figures/}}

\begin{document}

\begin{frontmatter}
  \title{High-resolution one-photon absorption spectroscopy of the $D\,{}^2\Sigma^- \leftarrow X\,{}^2\Pi$ system of radical OH and OD}

  \author{
    A.~N.~Heays,\textsuperscript{1}
    N.~de~Oliveira,\textsuperscript{2}
    B.~Gans,\textsuperscript{3}
    K.~Ito,\textsuperscript{2}
    S.~Boy\'e-P\'eronne,\textsuperscript{3}
    S.~Douin,\textsuperscript{3}
    K.~M.~Hickson,\textsuperscript{4,5}
    L.~Nahon,\textsuperscript{2}
    J.~C.~Loison\textsuperscript{4,5}
    \\[2ex]
    1. LERMA, Observatoire de Paris, PSL Research University, CNRS, Sorbonne Universit\'es, UPMC Univ. Paris 06, F-92190, Meudon, France \\
    2. Synchrotron SOLEIL, l’Orme des Merisiers, BP48 Saint Aubin, 91192 Gif sur Yvette Cedex, France\\
    3. Institut des Sciences Moléculaires d’Orsay (ISMO), CNRS UMR 8214, Univ Paris-Sud, Université Paris-Saclay, F-91405 Orsay, France \\
    4. Institut des Sciences Moléculaires (ISM), CNRS, Univ. Bordeaux, 351 cours de la Libération, 33400, Talence, France
    5. CNRS, Institut des Sciences Mol\'eculaires, UMR 5255, F-33400 Talence, France\\
  }
  \begin{abstract}
    Vacuum-ultraviolet (VUV) photoabsorption spectra  were recorded of the $A\,{}^2\Sigma^+(v'=0)\leftarrow{}X\,{}^2\Pi(v''=0)$, $D\,{}^2\Sigma^-(v'=0)\leftarrow{}X\,{}^2\Pi(v''=0)$ and $D\,{}^2\Sigma^-(v'=1)\leftarrow{}X\,{}^2\Pi(v''=0)$ bands of the OH and OD radicals generated in a plasma-discharge source using synchrotron radiation as a background continuum coupled with the VUV Fourier-transform spectrometer on the DESIRS beamline of synchrotron SOLEIL.
    High-resolution spectra permitted the quantification of transition frequencies, relative $f$-values, and natural line broadening.
    The $f$-values were absolutely calibrated with respect to a previous measurement of $A\,{}^2\Sigma^+(v'=0)\leftarrow{}X\,{}^2\Pi(v''=0)$ (\textcite{wang1979}).
    Lifetime broadening of the excited $D\,{}^2\Sigma^-(v=0)$ and $D\,{}^2\Sigma^-(v=1)$ levels is measured for the first time and compared with previous experimental limits, and implies a lifetime 5 times shorter than a theoretical prediction (\textcite{van_der_loo2005}).
    A local perturbation of the $D\,{}^2\Sigma^-(v=0)$ level in OH was found.
  \end{abstract}
\end{frontmatter}

\section{Introduction}

The OH radical plays an important role in atmospheric, interstellar, and plasma chemistry, and in combustion processes.
Laboratory work on the spectroscopy of neutral OH is made difficult by its transient nature requiring the use of  high-pressure electric discharges (e.g., Ref.~\citep{dilecce2012}), radio-frequency (RF) glow discharges \cite{bruggeman2012}, flames \cite{dieke1962,mercier1999,choudhuri2000}, laser photolysis \cite{mcraven2003,green2016}, or matrices \cite{cheng_bingming2001}.
It is frequently measured in the laboratory through the sensitive techniques of laser-induced fluorescence (via the well-known $A\,{}^2\Sigma^+ - X\,{}^2\Pi$ system) \cite{copeland1975,brzozowski1978} or by cavity ring-down spectroscopy \cite{spaanjaars1997,wang_chuji2009}. 

The ultraviolet (UV) and vacuum-ultraviolet (VUV) OH absorption systems, $A\,{}^2\Sigma^+ \leftarrow X\,{}^2\Pi$ and $D\,{}^2\Sigma^- \leftarrow X\,{}^2\Pi$, respectively, are detected in diffuse
interstellar clouds \cite{herbig1968,chaffee1977,weselak2009} and provide a valuable molecular tracer when rovibrational emission lines are lacking.
The photodissociation of OH by UV and VUV photons is also an important process in its chemical evolution and for related molecules in interstellar clouds, during star and planet formation (e.g., Refs.~\cite{van_dishoeck1984,snow2006,adamkovics2014}), in Earth and planetary atmospheres \cite{nair1994,moses2000}, and in cometary comae \cite{morgenthaler2007}.

The $A\,{}^2\Sigma^+(v') - X\,{}^2\Pi(v'')$ bands of OH and OD (sometimes abbreviated hereafter to $A(v')-X(v'')$) are spectroscopically well-characterised \cite{dieke1962,douglas1974,amiot1981,stark1994} and oscillator strengths ($f$-values) for the $A(0)\leftarrow X(0)$ absorption band are accurately known \cite{wang1979,wang1980}.
This has permitted this band to act as a tool for determining OH column densities in interstellar sight lines.

Absorption spectra of the $D\,{}^2\Sigma^-(v'=0) \leftarrow X\,{}^2\Pi(v''=0)$ band in OH and OD (abbreviated $D(0)\leftarrow X(0)$)  were recorded photographically in the 1970s \cite{douglas1974} utilising a discharge radical source, VUV \ce{H2} lamp, and grating spectrograph,  with a resolving power of \np{120000}.
Laser-based multi-photon absorption experiments have probed this and other Rydberg states of OH with high spectral-resolution using photoionisation detection schemes \cite{collard1991,de_beer1991,mcraven2003,green2016}.

An $f$-value for $D(0)\leftarrow X(0)$ in OH was deduced from interstellar-cloud absorption spectra \cite{chaffee1977} (also see the discussion of \textcite{roueff1996}). 
Laboratory studies measured the absolute cross sections of OH and OD VUV photoabsorption at low spectral resolution (with a resolving power $\lambda/\Delta\lambda =250$) \cite{nee1984,nee1984b} and roughly permitted another estimate of the $D(0)\leftarrow X(0)$ $f$-value for OH.
A series of theoretical studies \cite{langhoff1982,van_dishoeck1983,van_dishoeck1983b} calculated the photoabsorption and dissociation properties of many excited states of OH, including $D(0)$ and $D(1)$.
An accurate laboratory measurement of the $D(v')\leftarrow X(0)$ $f$-values would be useful to confirm the theoretical calculations and in future astronomical observations in the VUV.
This motivated the current experiment seeking an absolute $D(0)\leftarrow X(0)$ $f$-value.
The $A(0)\leftarrow X(0)$ band was observed again here to provide a reference standard in the absence of a direct column density calibration.

Lifetimes of bound levels of the $D\,{}^2\Sigma^-$ state are limited by radiative decay to the ground or lower-lying excited states and by nonradiative predissociation.
These processes have been studied in detail theoretically \cite{van_dishoeck1983b,van_der_loo2005} and laboratory-measured lifetimes (or, inversely, transition linewidths) have provided an indication of lower and upper limits \cite{de_beer1991,mcraven2003}.
Measuring this line broadening due to short $D(0)$ and $D(1)$ lifetimes is a further goal of this experiment.

In what follows, we describe rotational levels of the $A\,{}^2\Sigma^+$ and $D\,{}^2\Sigma^-$ states by alternative quantum numbers $J$ or $N$, where the latter neglects the spin angular momentum of OH's unpaired electron.
$J$ is the more rigorous quantum number in the presence of spin-orbit interaction but the weakness of this interaction means that patterns of level energies and lifetimes more closely follow $N$ \cite{herzberg1989}.
Single and double primes indicate excited and ground state levels, for example, $N'$ and $N''$ respectively.
We describe the substates of each electronic state according to the conventional notation regarding $F_i$ numbering:
\begin{align*}
  X\,{}^2\Pi &
               \begin{cases}
                 F_1,\ e/f\textrm{ parity},\ J=N+1/2\\
                 F_2,\ e/f\textrm{ parity},\ J=N-1/2\\
               \end{cases} \\
  A\,{}^2\Sigma^+ &
                    \begin{cases}
                      F_1,\ e\textrm{ parity},\ J=N+1/2\\
                      F_2,\ f\textrm{ parity},\ J=N-1/2\\
                    \end{cases} \\
  D\,{}^2\Sigma^- &
                    \begin{cases}
                      F_1,\ f\textrm{ parity},\ J=N+1/2\\
                      F_2,\ e\textrm{ parity},\ J=N-1/2\\
                    \end{cases} \\
\end{align*}
Rotational transitions are labelled accordingly so that, for example, $Q_{21fe}(J'')$ indicates a $J'=J''$ transition from the $F_1$ $e$-parity ground state substate to a $F_2$ $f$-parity excited state substate.

\section{Experimental description}

The DESIRS beamline at the SOLEIL synchrotron is equipped with an undulator producing radiation in the UV-VUV spectral range with a bandwidth of $\Delta \nu / \nu = 7\%$ acting as a background emission continuum, where $\nu$ is the photon frequency or wavenumber.
High-resolution broadband absorption spectra of the full bandwidth are recorded when this is directed towards the Fourier-transform spectrometer (FTS) branch of the beamline.
A large multipurpose sample chamber is located upstream of the FTS and is equipped with a double-stage differential pumping system, allowing for the easy installation of various absorption cell configurations within.
The beamline and FTS photoabsorption branch have been previously described in detail \cite{de_oliveira2011,de_oliveira2016}.
Here, we provide details of the elements relevant to the present work. 

Two methods were used to generate OH and OD radicals from sample gases of \ce{H2O} and \ce{D2O} in three separate campaigns, hereafter identified as DC\,I,  DC\,II, and RF.

A direct-current (DC) electric discharge inside a windowed cell was constructed and operated in vacuum inside the multipurpose sample chamber.
The cell is a glass tube of 300\,mm length and 10\,mm inner diameter enclosed at both ends by 1\,mm thick \ce{MgF2} windows.
The cathode is a T-shaped stainless steel DN16 tube connected to the inside of the cell, and a stainless steel anode is mounted onto a cross-shaped glass holder at its opposite end.
The distance between cathode and anode is approximately 250\,mm.
All parts of the cell are refrigerated by a water circuit  leading to a chiller outside the vacuum chamber.
The temperature of the cell was monitored with a thermocouple in order to avoid any in-vacuum damage during operation.
The cell temperature constantly increased during the DC\,I campaign, despite the presence of water cooling, and required periods for system cool-down and a slowing of the data acquisition rate.
The design of the cooling system was improved for the DC\,II experiment and this permitted continuous operation, more efficient data collection, and better quality spectra.
The current in the discharge was respectively set to 60 and 50\,mA in the DC\,I and DC\,II experiments, and the precursor water vapour was diluted in helium and introduced into the system through a needle valve.
The cell pressure was selected after consideration of trial absorption spectrum, as discussed further below.

We also operated a radio-frequency (RF) discharge cell, with its installation requiring a complete removal of the standard sample chamber. 
The discharge cell is a borosilicate glass tube sealed at both ends by 1\,mm-thick \ce{MgF2} windows located 1500\,mm apart.
The tube inner diameter is 16\,mm\ and the whole system is sealed with Viton O-rings.
The synchrotron beamline and FTS chamber were connected to the discharge cell by stainless steel 6-way CF40 cross tubes.
These were pumped by 300\,L\,s$^{-1}$ turbo molecular pumps in order to maintain the approximately \np[mbar]{5e-8} vacuum required by the beamline and FTS chamber.

The RF discharge cell was filled with a continuous flow of \ce{H2O} or \ce{D2O} vapour and He carrier gas, and generated a capacitively coupled air-cooled RF plasma operating at 13.56\,MHz with a variable input power between 0 and 600\,W.
The RF-generated plasma showed a visible path length of 1200\,mm with the input power was set to 150\,W.
The cell is installed within the helical coil antenna of the RF source.
A reservoir of water was connected to the cell and its vapour pressure controlled with a leak valve.
The He was introduced to the cell through apertures adjacent to both windows in both RF and DC experiments in order to prevent their contamination by discharge products.
The cell was continuously pumped by a 600\,m$^{3}$\,h$^{-1}$ roots pump. 

The $D(v')\leftarrow X(0)$ and $A(0)\leftarrow X(0)$ bands are too far apart to appear in a single undulator spectral window.
A spectrum containing $D(0)\leftarrow X(0)$ and $D(1)\leftarrow X(0)$ was first measured in the VUV immediately followed by the recording of $A(0)\leftarrow X(0)$ in the UV without altering the discharge settings.
This permitted the measurement of their relative $f$-values.
Actually, the OH column density was observed to vary within 10\% during each experiment.
We then adopt a systematic uncertainty of $\pm5$\% for all measured band strengths.
The $f$-value ratio of bands measured from two different measurements then incurs an uncertainty of 7\%.

We selected a spectral resolution to obtain an acceptable absorption-to-noise ratio under each set of experimental conditions, that is, 0.27, 0.54, and 1.1\,cm$^{-1}$\,FWHM in experiments RF, DC\,I, and DC\,II; respectively.
A typical integration time for recording the UV $A(0)\leftarrow X(0)$ band was 15\,min, and 130\,min was required to record the VUV spectral window.
This increase is due to the proximity of the $D(0)\leftarrow X(0)$ and $D(1)\leftarrow X(0)$ bands to the short-wavelength transmission limit of the \ce{MgF2} windows that resulted in a sharp attenuation of the synchrotron beam.
In addition, residual water vapour inside the discharge shows increasing absorption at shorter wavelengths.
The signal-to-noise ratio in the optical depth of a 50\%-absorbed line at 122\,nm was 45:1 after optimising the composition and conditions in the RF discharge.

Residual water in the discharge strongly absorbs in the VUV where the $D\leftarrow X$ bands are found.
The density and dilution of precursor water was optimised to maintain a sufficient level of transmission in this region while retaining a large-enough column of OH/OD to permit observation of the 10-times weaker $A(0)\leftarrow X(0)$ band.
This optimisation was monitored using the strong emission of the 308\,nm $A\rightarrow X$ OH band by means of a commercial UV-visible spectrometer.

An absolute wavenumber calibration of the $A(0)\leftarrow X(0)$ lines appearing in our OH and OD spectra was made with reference to the measurements of \textcite{stark1994}, who claim an estimated frequency uncertainty of \np[cm^{-1}]{0.0005}.
This also served to calibrate the $D\leftarrow X$ spectra with the help of an additional frequency standard internal to the FTS arising from a secondary laser interferometer controlling its mirror translation.
The experimental wavenumbers we find below have a random uncertainty of the order of \np[cm^{-1}]{0.01}.

High-resolution spectra of \ce{H2O} and \ce{D2O} were recorded under similar conditions to the discharge experiment to help decipher precursor and radical signals.
The column density of water molecules was monitored after ignition of the various discharges and this indicated that approximately 50\% are dissociated.

\section{Analysis method}

Previous studies \cite{dieke1962,amiot1981,bernath2009,coxon1980,douglas1974,collard1991,de_beer1991,stark1994} have established the rotational energy levels of the $X(0)$, $A(0)$ and $D(0)$ states of OH and OD and the $D(1)$ state of OH, and fit these to minimally-specified effective Hamiltonians.
The lack of abrupt perturbations afflicting these levels permitted these representations to accurately model the transition energies of an entire absorption band with only a few parameters.
We chose to fit simulated bands directly to our experimental spectra rather than measuring line positions and intensities individually.
This allowed for the accurate fitting of those bands where individual rotational lines have poor signal-to-noise ratios, and in one case permitted the identification of a weak local perturbation.

We adopted an effective Hamiltonian for the $X\,{}^2\Pi$ state according to the formulation of \textcite{brown1979}.
Matrix elements for this are conveniently tabulated by \textcite{amiot1981} (see their Table~II), who also provide an appropriate Hamiltonian for levels of both the $A\,{}^2\Sigma^+$ and $D\,{}^2\Sigma^-$ states (see their Table~III).
For the better-known $X(0)$ and $A(0)$ levels of OH and OD we did not attempt to determine new molecular parameters from our measured spectra, instead employing reference data.
For $D(0)$ and $D(1)$ we adjusted the various effective Hamiltonian parameters to best fit the experiment.

The line strength, $S$, of a particular rotational transition within an $A-X$ or $D-X$ band may be written as the product of two factors,
\begin{equation}
  S=S_{v'v''}\times\mathcal{S}.
  \label{eq:line strengths}
\end{equation}
The former term is the band strength, $S_{v'v''}$, comprising an electronic transition moment and vibrational overlap factor, and is fit to our experimental spectra and reference data as a single value for each band.
The rotational line strength factors, $\mathcal{S}$, are specific to each branch and rotational transition, and are affected by the spin-rotation mixing of Hund's case (a) ground state $F_1$ and $F_2$ levels with common parity.
A good reference for the calculation of rotational line strength factors when such mixing occurs is that of \textcite{hougen1970}.
Effective Hamiltonians describing the OH and OD ground states and their spin-rotation mixing are well known from previous studies and provide the necessary rotational line strength factors for the modelling of our experimental spectra.

The relations between line strength, line $f$-value ($f$) and integrated line cross section ($\sigma^\text{int}$) are \cite{larsson1983,bernath_book2005}
\begin{equation}
  f = \frac{\np{3.038e-6}\nu S}{2J''+1}
  \label{eq:fvalue}
\end{equation}
and
\begin{equation}
  \sigma^\text{int} = \frac{f\alpha}{\np{1.1296e12}}
\end{equation}
where $\nu$ is the transition wavenumber, $\alpha$ is the fractional population of molecules in the ground state level, and the listed constants assume the following units: $S[\text{au}]$, $\nu[\text{cm}^{-1}]$, $\sigma^\text{int}[\text{cm}^2\text{cm}^{-1}]$.
We assume a Boltzmann distribution of ground state rotational levels with the associated temperature fit to best match the experimental spectra.

A band $f$-value, $f_{v'v''}$, may also be factored out of the line $f$-value according to \cite{larsson1983}
\begin{equation}
  f_{v'v''} = \frac{2J''+1}{2}\frac{f}{\mathcal{S}}
\end{equation}
for a ${}^2\Pi-{}^2\Sigma^\pm$ transition, and these are approximately constant for all lines in each observed band.
A small variation of $f_{v'v''}$ with molecular rotation is known to occur for the $A-X$ bands of OH and OD \cite{wang1979,luque1998} due to the relevant electronic transition moment possessing some internuclear-distance dependence.

To simulate our experimental spectra, transition energies and line strengths were calculated from effective Hamiltonians and an assumed band strength.
These line strengths were converted into line-integrated cross sections assuming a rotational temperature and then into a wavelength-dependent cross section, $\sigma(\nu)$, by assuming a Voigt lineshape for each rotational line.
The Gaussian width component of this lineshape corresponds to  Doppler broadening (fixed to the rotational temperature).
Its Lorentzian component is attributable to lifetime broadening and was fit to the spectra where required.
The sum of line cross sections was converted into a transmission spectrum, $I(\nu)$, by application of the Beer-Lambert law,
\begin{equation}
  I(\nu) = I_0(\nu)\exp\biggl(-N\sum\sigma(\nu)\biggr).
\end{equation}
Here, $I_0(\nu)$ is the wavelength-dependent intensity of the synchrotron beam (observable in the windows between OH and \ce{H2O} absorption lines) and $N$ is the column density of OH molecules in the absorption cell, which we calibrated to the known strength of the OH $A(0)-X(0)$ band.
The model spectrum was convolved with the sinc-shaped instrumental broadening inherent to the FTS instrument (which is also responsible for apparent spurious emission lines on either side of deep and narrow absorption lines in the experimental spectra).
Finally, a pointwise comparison was made between model and recorded spectra and the various model parameters automatically adjusted until optimal agreement was achieved using a least-squares algorithm.

Additional features were added to the model spectrum to better reproduce the experiment.
There was significant absorption from the precursors \ce{H2O} or \ce{D2O} which was accounted for by including cross sections for these species measured with the discharge off.
Additional lines of \ce{H2} and \ce{D2} were also accounted for, some of which showed significant ground-state excitation with $J''$ as high as 7 and $v''$ up to 3.
Other features appear at the same frequency in both \ce{H2O} and \ce{D2O} discharges.
These are then due to molecular oxygen formed in the discharge or other contaminants.
In particular, a strong absorption feature of metastable \ce{O2} at \np[cm^{-1}]{80400} \cite{ogawa1975} is evident in our spectrum (not shown) and also likely contributes to the 124\,nm peak observed in the OH spectrum recorded by \citet{nee1984}.

\begin{figure*}
  \centering
  \includegraphics{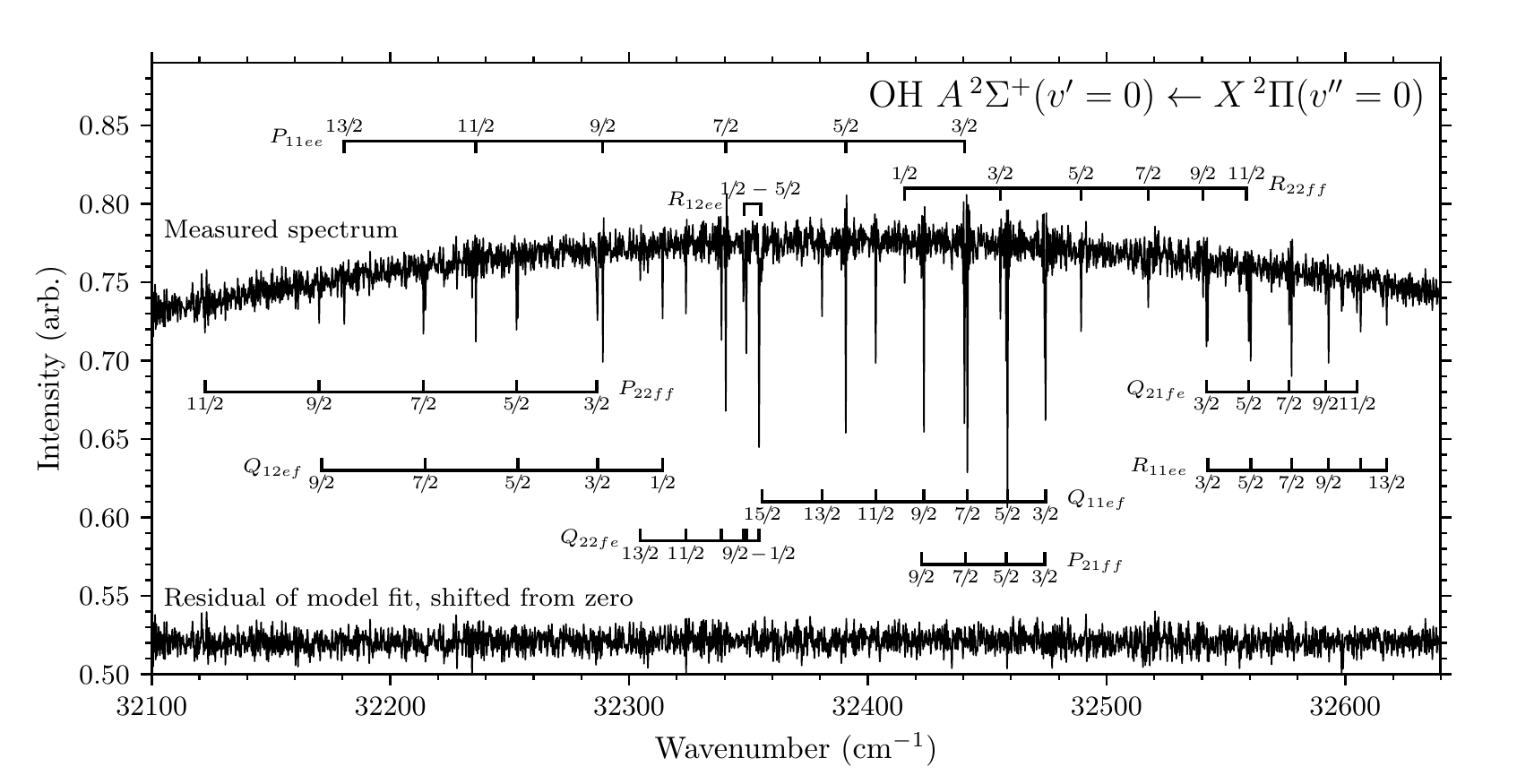}
  \caption{
    Assigned experimental spectrum (RF discharge, \np[cm^{-1}\,FWHM]{0.27} resolution) showing OH $A\,{}^2\Sigma^+(v'=0)\leftarrow X\,{}^2\Pi(v''=0)$ and the residual fit of a band model.
    Rotational transitions are enumerated by $J''$.
    The pseudo-Gaussian profile of the undulator source causes a slow variation of UV intensity.
    }
  \label{fig:OH A(0)-X(0) spectrum}
\end{figure*}

\section{Results}

\subsection{OH $A\,{}^2\Sigma^+(v'=0)\leftarrow X\,{}^2\Pi(v''=0)$}
\label{sec:OH A(0)}

Molecular parameters for the $X(0)$ state of OH were generated by fitting an effective Hamiltonian to the term values given by \textcite{coxon1980} and used without modification in the subsequent analysis of OH bands.
Molecular parameters for the $A(0)$ state of OH were generated from the term values found to high precision by \textcite{stark1994}.
An $A(0)\leftarrow X(0)$ model spectrum was then generated by combining these two Hamiltonians unaltered and determining a best-fit ground state rotational temperature and band $f$-value.
This provided a perfect fit to the experimental spectrum recorded in the RF discharge, shown in Fig.~\ref{fig:OH A(0)-X(0) spectrum}, along with a modelled spectrum and residual difference.

The best-fit rotational temperature of the $A(0)\leftarrow X(0)$ spectrum is $395\pm7$\,K and was assumed to remain constant and stable for the subsequent analyses of $D(0)\leftarrow X(0)$ and $D(1)\leftarrow X(0)$ bands recorded with the RF discharge.
Two other $A(0)\leftarrow X(0)$ spectra were recorded with lower spectral resolution in the DC discharge cell and with similar OH rotational temperatures.
The natural line broadening of $A(0)\leftarrow X(0)$ is significantly smaller than our experimental resolution due to its purely-radiative decay and long lifetime (about \np[ns]{800} \cite{brzozowski1978}).

\subsection{OH $D\,{}^2\Sigma^-(v'=0)\leftarrow X\,{}^2\Pi(v''=0)$}
\label{sec:OH D(0)}

\begin{figure*}
  \centering
  \includegraphics{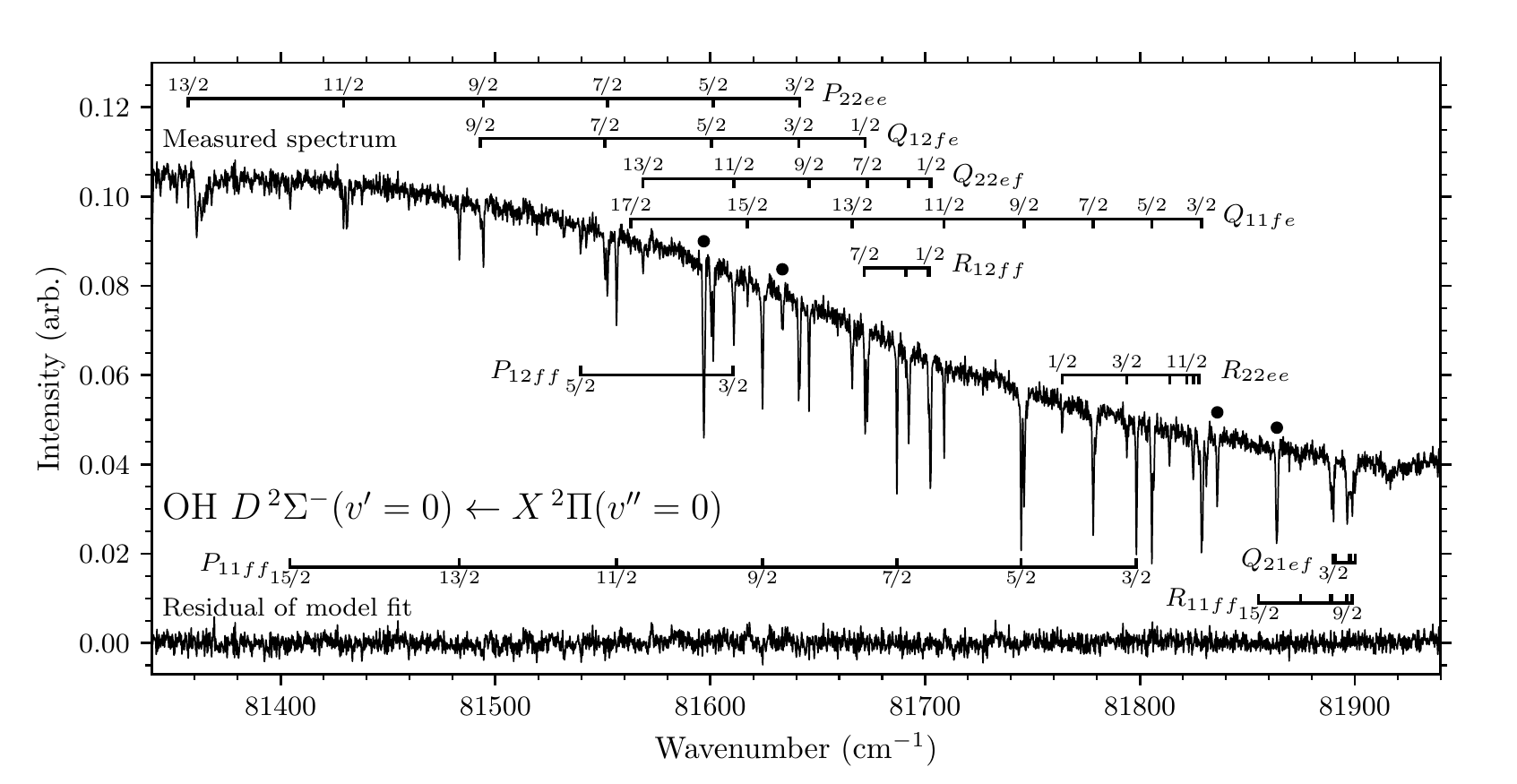}
  \caption{
    Assigned experimental spectrum (RF discharge, \np[cm^{-1}\,FWHM]{0.27} resolution) showing the OH $D\,{}^2\Sigma^-(v'=0)\leftarrow X\,{}^2\Pi(v''=0)$ band and the residual fit of a band model.
    Rotational transitions are enumerated by $J''$.
    \ce{H2} absorption lines are indicated with circles.
    The asymmetric absorption visible at \np[cm^{-1}]{81360} is due to metastable \ce{O2} \protect\cite{ogawa1975}.
    A smooth decrease of intensity with wavenumber is due to \ce{H2O} continuum absorption.
  }
  \label{fig:OH D(0)-X(0) spectrum}
\end{figure*}
The experimental spectrum of the OH $D(0)\leftarrow X(0)$ band shown in Fig.~\ref{fig:OH D(0)-X(0) spectrum} was fit initially to the RF spectrum with a band model adopting an effective Hamiltonian describing the $D(0)$ level.
This fit proved unsatisfactory, with some model transitions shifted from their observed frequencies by up to \np[cm^{-1}]{0.15}.
A subsequent fit was then made where each $D(0)$ energy level up to $N=8$ was independently varied, producing the residual fit plotted in Fig.~\ref{fig:OH D(0)-X(0) spectrum}.
\begin{figure}
  \centering
  \includegraphics{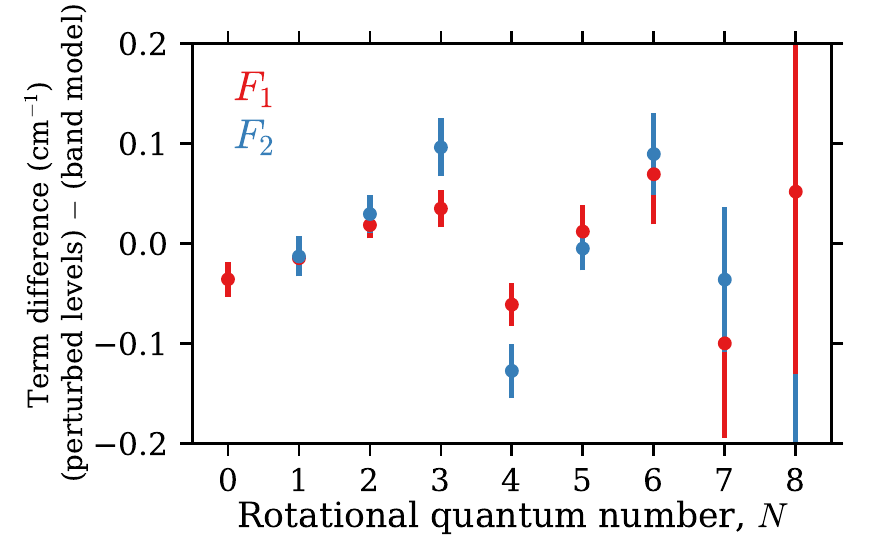}
  \caption{Difference between OH $D\,{}^2\Sigma^-(v=0)$ rotational energy levels fitted to the experimental spectrum assuming an effective Hamiltonian band model and fitted individually. The observed differences are due to the occurrence of a perturbation not included in the effective Hamiltonian.   Error bars refer to the ($1\sigma$) estimated uncertainty of individually-fitted level energies.}
  \label{fig:D(0) term values}
\end{figure}
The differences between band-modelled and individually-fitted level energies are plotted in Fig.~\ref{fig:D(0) term values} and show a characteristic pattern between $N=3$ and 4 whereby the rotational series of an unknown state is crossing $D(0)$. 

\begin{figure}
  \centering
  \includegraphics{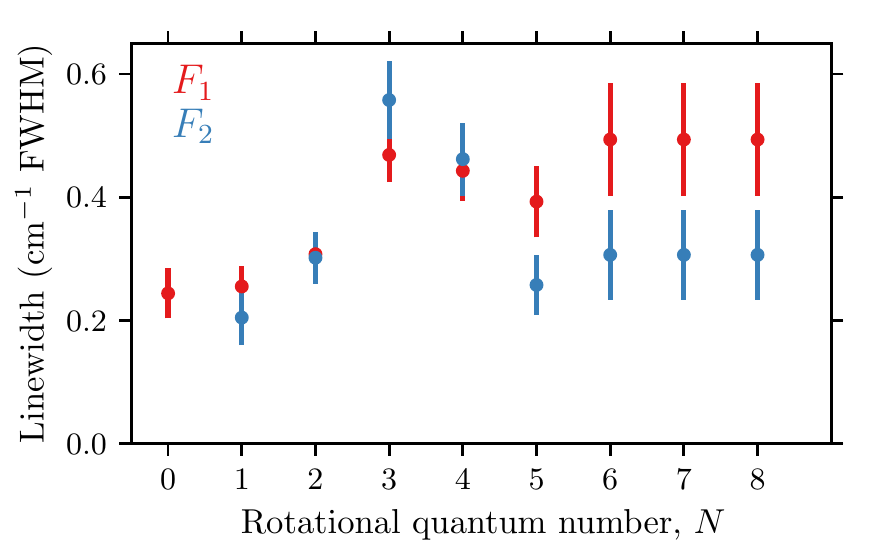}
  \caption{Linewidths of OH $D\,{}^2\Sigma^-(v=0)$ levels with estimated ($1\sigma$) fitting uncertainties.  Widths for $F_1$ and $F_2$ levels with $N\geq 6$ were fit to common values.}
  \label{fig:D(0) linewidths}
\end{figure}
To accurately model the experimental spectrum it was also necessary to introduce line broadening, treated as $N'$-dependent linewidths.
These are shown in Fig.~\ref{fig:D(0) linewidths} to increase with $N'$, and with a maximum value near $N'=3$ and 4 of about \np[cm^{-1}\,FWHM]{0.5}, with a corresponding minimum lifetime of about \np[ps]{10}.
The coincidence of this maximum with the maximum level-energy perturbation is discussed further in Sec.~\ref{sec: OH D(0) perturbation}.

The $N$-dependent linewidth and level shifts determined from the RF spectrum were verified by the lower-resolution DC\,I and DC\,II spectra.

\subsection{OH $D\,{}^2\Sigma^-(v'=1)\leftarrow X\,{}^2\Pi(v''=0)$}

\begin{figure*}
  \centering
  \includegraphics{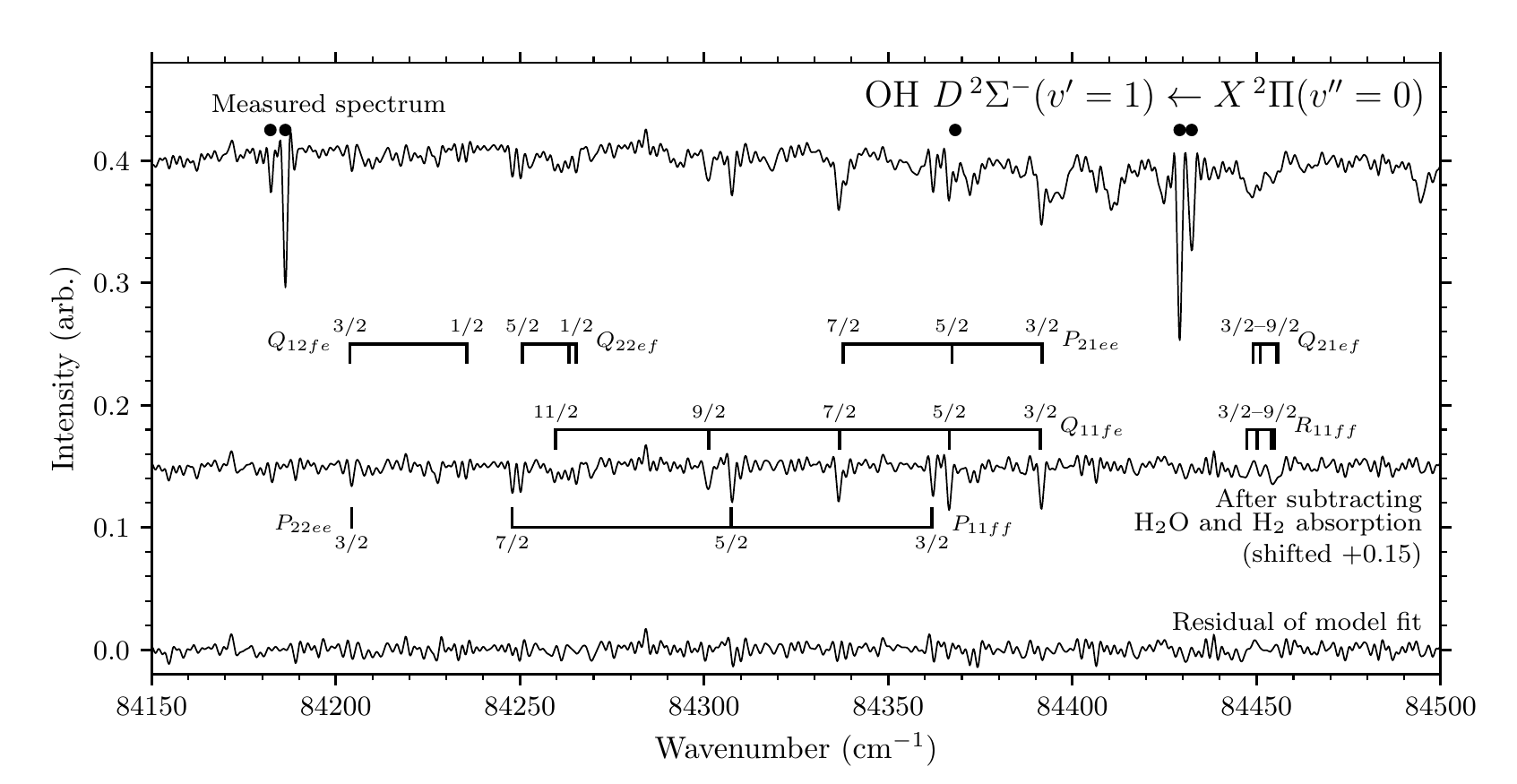}
  \caption{
    Assigned experimental spectrum (DC\,I discharge, \np[cm^{-1}\,FWHM]{1.1} resolution) showing the OH $D\,{}^2\Sigma^-(v'=1)\leftarrow X\,{}^2\Pi(v''=0)$ band and the residual fit of a band model.
    Rotational transitions are enumerated by $J''$.
    \ce{H2} absorption lines are indicated with circles.
    An additional trace illustrates the presence of \ce{H2O} and \ce{H2} contaminant lines.
  }
  \label{fig:OH D(1)-X(0) spectrum}
\end{figure*}

Absorption transitions to OH $D(1)$ levels with $N'\leq 5$ were observable in the DC\,I spectrum after subtracting the structured absorption due to \ce{H2O}, as shown in Fig.~\ref{fig:OH D(1)-X(0) spectrum}.
A model of this spectrum was constructed assuming the $D(1)$ molecular constants deduced by \textcite{de_beer1991}.
The final agreement between experimental and model bands is also shown in Fig.~\ref{fig:OH D(1)-X(0) spectrum} and required the assumption of a rotation-independent linewidth of $0.25\pm0.10$\,cm$^{-1}$\,FWHM.

\subsection{OD $A\,{}^2\Sigma^+(v'=0)\leftarrow X\,{}^2\Pi(v''=0)$}

\begin{figure*}
  \centering
  \includegraphics{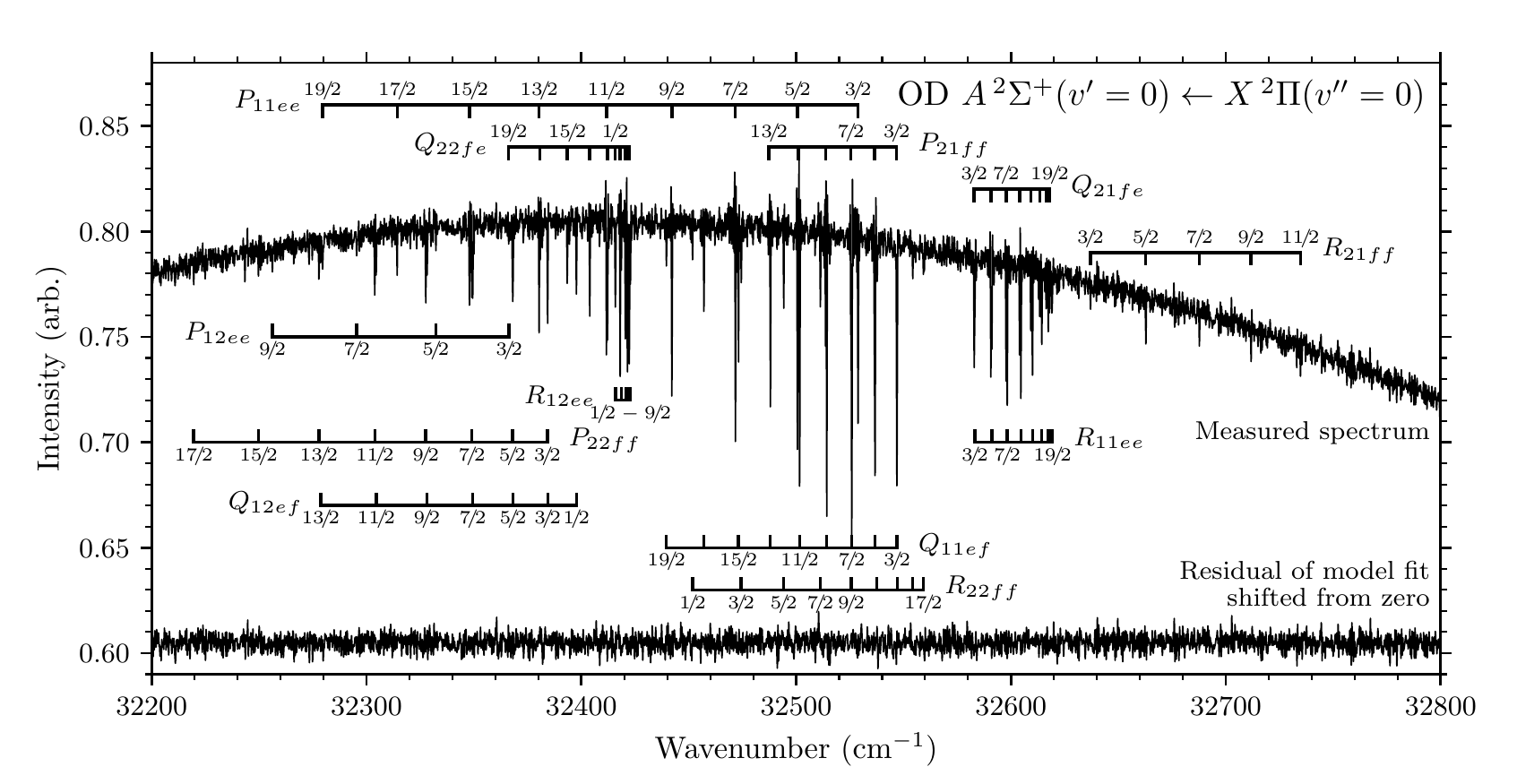}
  \caption{
    Assigned experimental spectrum (RF discharge \np[cm^{-1}\,FWHM]{0.27} resolution) showing the OD $A\,{}^2\Sigma^+(v'=0)\leftarrow X\,{}^2\Pi(v''=0)$ band and the residual fit of a band model.
    Rotational transitions are enumerated by $J''$.
    The pseudo-Gaussian profile of the undulator source causes a slow variation of UV intensity.
  }
  \label{fig:OD A(0)-X(0) spectrum}
\end{figure*}

Effective Hamiltonians were generated to describe the $X(0)$ and $A(0)$ states of OD based on the experimental term values of \citet{stark1994}.
These were then used to model the OD $A(0)\leftarrow X(0)$ RF spectrum shown in Fig.~\ref{fig:OD A(0)-X(0) spectrum}, leading to a best-fit rotational temperature of $363\pm4$\,K.

\subsection{OD $D\,{}^2\Sigma^-(v'=0)\leftarrow X\,{}^2\Pi(v''=0)$}

\begin{figure*}
  \centering
  \includegraphics{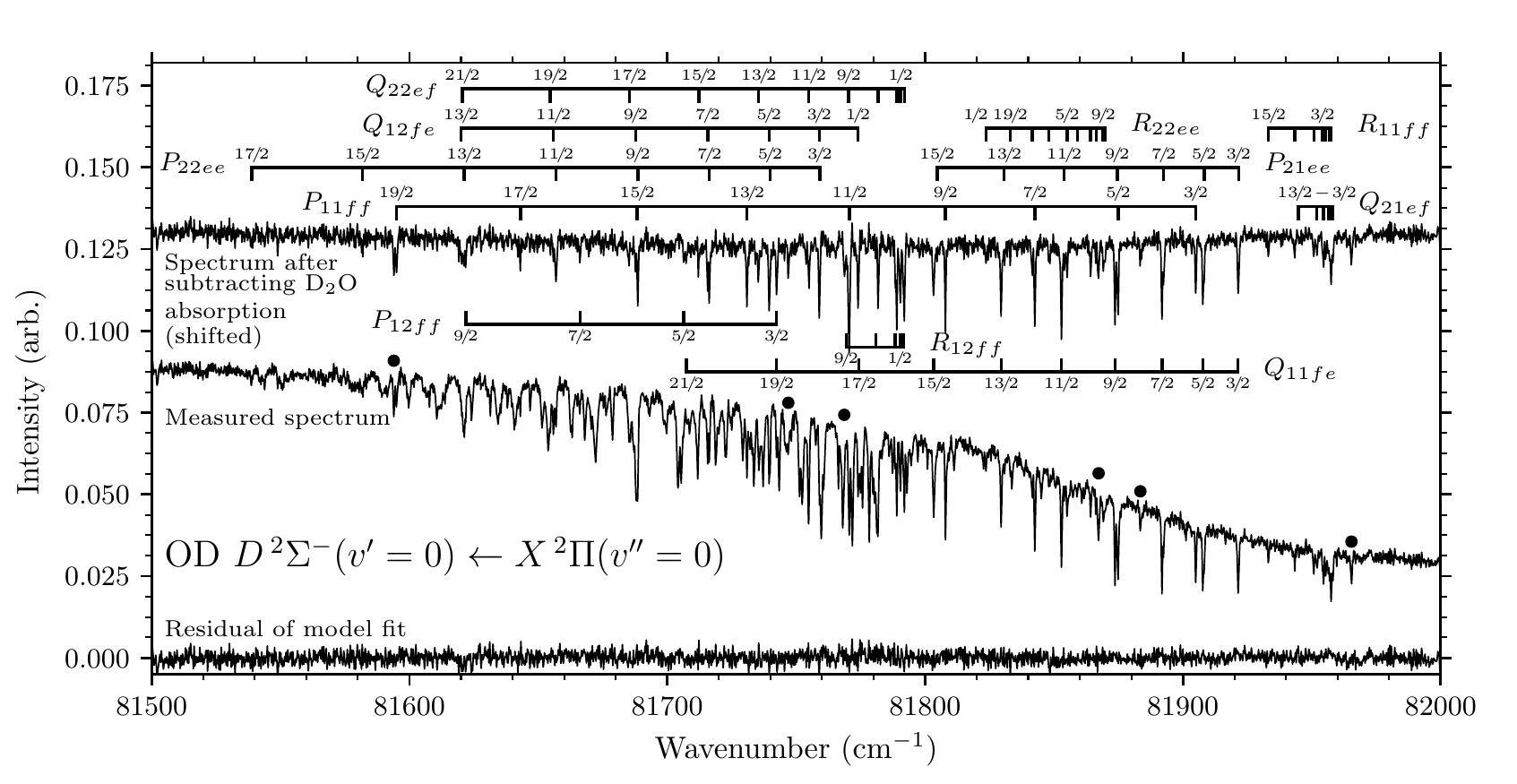}
  \caption{
    Assigned experimental spectrum (RF discharge, \np[cm^{-1}\,FWHM]{0.27} resolution) showing the OD $D\,{}^2\Sigma^-(v'=0)\leftarrow X\,{}^2\Pi(v''=0)$ band and the residual fit of a band model.
    Rotational transitions are enumerated by $J''$.
    \ce{D2} absorption lines are indicated with circles.
    A further trace indicates the significant contamination by structured and continuum \ce{D2O} absorption.
  }
  \label{fig:OD D(0)-X(0) spectrum}
\end{figure*}
The experimental RF spectrum of OD $D(0)\leftarrow X(0)$ is shown in Fig.~\ref{fig:OD D(0)-X(0) spectrum}.
The unfortunate blending with many sharp features of \ce{D2O} required the subtraction of a reference spectrum with the RF discharge turned off.
A constant natural line broadening of $0.25\pm0.05$\,cm$^{-1}$ was required to satisfactorily fit the observed spectrum, and no local perturbation like the one affecting $D(0)$ in OH was observed.

\subsection{OD $D\,{}^2\Sigma^-(v'=1)\leftarrow X\,{}^2\Pi(v''=0)$}

\begin{figure*}
  \centering
  \includegraphics{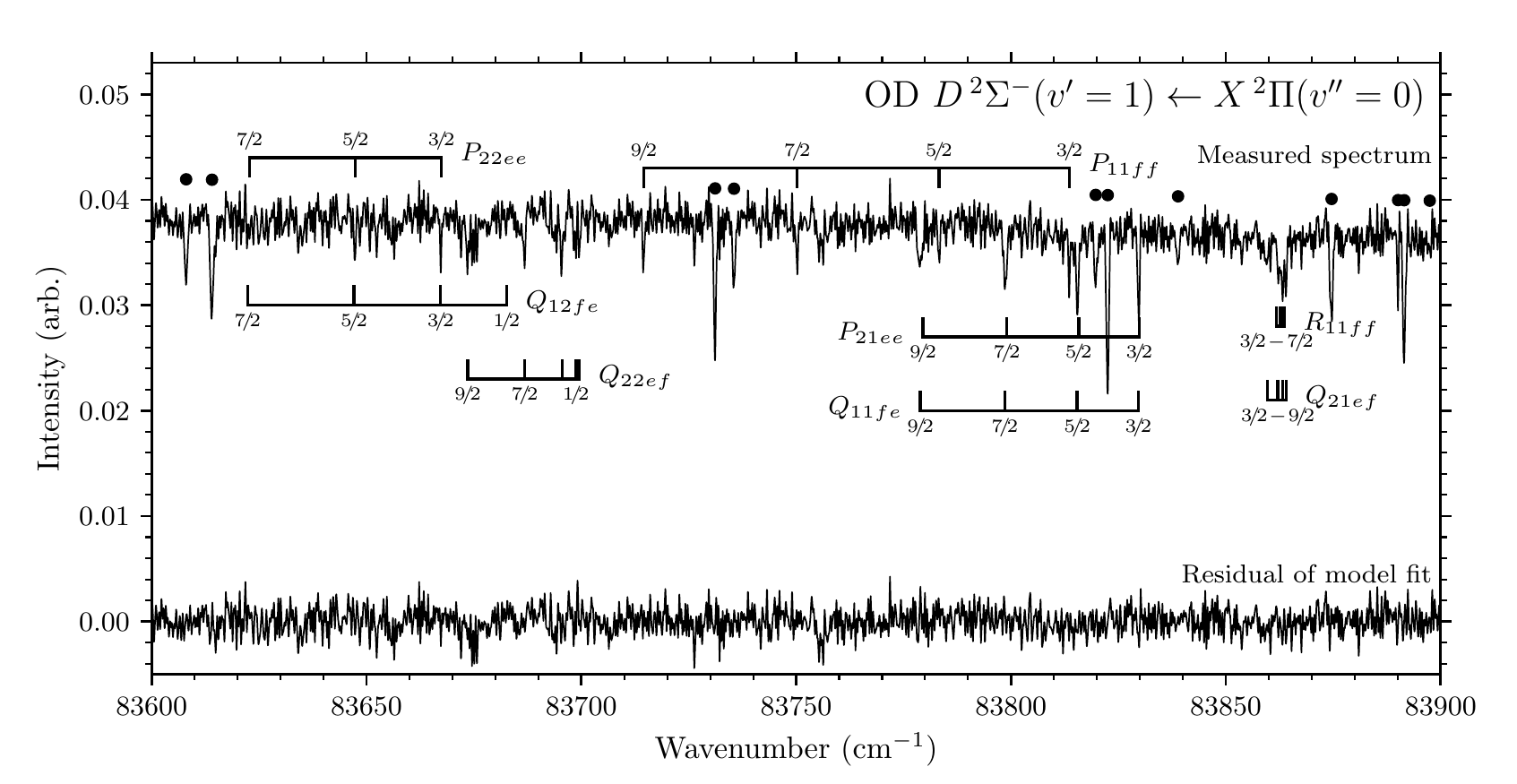}
  \caption{
    Assigned experimental spectrum (RF discharge, \np[cm^{-1}\,FWHM]{0.27} resolution) showing the OD $D\,{}^2\Sigma^-(v'=1)\leftarrow X\,{}^2\Pi(v''=0)$ band and the residual fit of a band model.
    Rotational transitions are enumerated by $J''$.
    \ce{D2} absorption lines are indicated with circles.
  }
  \label{fig:OD D(1)-X(0) spectrum}
\end{figure*}
Absorption due to $D(1)\leftarrow X(0)$ is shown in Fig.~\ref{fig:OD D(1)-X(0) spectrum}.
A model was fitted to this spectrum, providing the molecular parameters listed in Table~\ref{tab:Hamiltonian constants}.
Only a few rotational transitions are significantly above the noise level, but these were sufficient to deduce a relative band $f$-value for this band and a common linewidth for all transitions of $0.29\pm0.10$\,cm$^{-1}$.

\subsection{Term values and effective Hamiltonians}

\begin{table*}
  \centering
  \begin{minipage}{\linewidth}
    \centering
    \caption{Molecular parameters fit to the experimental spectra or reference data (cm$^{-1}$).%
      \protect\footnote{The listed constants correspond to the ${}^2\Pi$ ($X$ state) and ${}^2\Sigma^{\pm}$ ($A$ and $D$ states) effective Hamiltonians of \textcite{amiot1981} (Tables II and III). Constants that are not listed were fixed to zero.
      The estimated $1\sigma$ uncertainty of parameters fitted to best reproduce the experimental spectra are given in parentheses in units of its least significant digit. These are fitting errors only and do not account for possible model error in contaminated spectra.  Parameters without an uncertainty were determined from reference data and held fixed while fitting the spectra.
      Absolute term values are calculated with respect to the transition wavenumbers in \textcite{stark1994} and include the 0.0005\,cm$^{-1}$ systematic calibration uncertainty of this reference.
      These constants are constrained by level energies with $N\lesssim 10$ (or as indicated) and will provide a limited-accuracy extrapolation above this limit.
    }%
    }%
    
    \label{tab:Hamiltonian constants}
    \footnotesize
    \begin{tabular}{c}
      \toprule
      {\large OH $X\,{}^2\Pi(v=0)$\footnote{Fitted to the term values deduced by \textcite{coxon1980}.}}     \\
      $T        = \np{38.2183}          $ \qquad
      $A        = \np{-139.0746}         $ \qquad
      $B        = \np{18.5504}          $ \\
      $D        = \np{0.001918}       $ \qquad
      $H        = \np{1.41e-7}      $ \qquad
      $L        = \np{-1.11e-11}      $ \\
      $p        = \np{0.1965}         $ \qquad
      $p_D      = \np{-7.11e-6}     $\qquad
      $p_H       = \np{-7.69e-9}     $ \\
      $q        = \np{-0.0388}       $ \qquad
      $q_D      = \np{1.465e-5}      $ \qquad
      $q_H       = \np{-2.38e-9}     $ \\
      $\gamma   = \np{-0.0222}       $ \qquad
      $\gamma_D = \np{-2.55e-6}      $ \qquad
      $\gamma_H = \np{-6.07e-6}     $ \\
      \midrule
      {\large OH $A\,{}^2\Sigma^+(v=0)$\footnote{Fitted to the term values deduced by \textcite{stark1994}.}} \\
      $T        = \np{32440.5730}$ \qquad
      $B        = \np{16.9652}$ \\
      $D        = \np{2.06e-3}$ \qquad
      $H        = \np{1.28e-7}$ \qquad
      $L        = \np{2.0e-11}$ \\
      $\gamma   = \np{0.226 }$ \qquad
      $\gamma_D = \np{-4.60e-5}$ \qquad
      $\gamma_H = \np{2.3e-9}$ \\
      \midrule
      {\large OH $D\,{}^2\Sigma^-(v=0)$\footnote{These constants do not reproduce the perturbed spectrum within experimental accuracy, see Sec.~\ref{sec:OH D(0)} and Table~\ref{tab:OH_D00_terms}.}} \\
      $T        =  81798.3457(84)     $ \qquad  % new Nelson data 2017-05-26
      $B        =  15.201(14)         $ \\
      $D        =  0.001384(63)       $ \qquad
      $H        = -2.80(69)\times10^{-8}$ \\
      $\gamma   = -0.3538(48)         $ \qquad
      $\gamma_D =  0.00254(38)        $ \\
      $\gamma_H = -3.53(62)\times10^{-5}$ \\
      \midrule
      {\large OH $D\,{}^2\Sigma^-(v=1)$\footnote{$T$ was adjusted to best match the measured spectra, other constants were taken from \textcite{de_beer1991}.}}     \\
      $T = 84361.745(33)   $ \qquad %calibrated to stark1994 using 160909 D00 as an intermediary
      $B   =      14.79  $ \qquad
      $D   =   0.00172  $ \qquad
      $\gamma    =   -0.3  $ \\
      % \midrule
      \bottomrule
      \\[\medskipamount]
      \toprule
      {\large OD $X\,{}^2\Pi(v=0)$\textsuperscript{\emph{c}}} \\
      $T   =\np{51.6031}   $ \qquad
      $A   =\np{-139.0238} $ \qquad
      $A_D =\np{-8.80e-4}  $ \\
      $A_H =\np{3.82e-7}   $ \qquad
      $B   =\np{9.8831}    $ \qquad 
      $D   =\np{5.391e-4}  $ \\ 
      $H   =\np{2.15e-8}   $ \qquad 
      $L   =\np{6.29e-13}  $ \qquad 
      $p   =\np{0.114}     $ \\ 
      $q   =\np{-0.0109}   $ \qquad 
      $q_D =\np{1.64e-6} $ \\ 
      \midrule
      {\large OD $A\,{}^2\Sigma^+(v=0)$\textsuperscript{\emph{c}}} \\
      $T        = 32528.873 $ \qquad
      $B        = \np{9.0437}       $ \qquad
      $D        = \np{5.78e-4}      $ \\
      $H        = \np{1.80e-8}      $ \qquad
      $\gamma   = \np{0.121}        $\qquad
      $\gamma_D = \np{-1.33e-5}     $\\
      \midrule
      {\large OD $D\,{}^2\Sigma^-(v=0)$}     \\
      $T     = 81904.8700(92)$\qquad    % new Nelson data 2017-05-26 calibrated down 0.06
      $B     = 8.22893(45)   $      \\
      $D     = 0.0005475(55) $\qquad
      $\gamma= -0.1810(15)   $\\
      \midrule
      {\large OD $D\,{}^2\Sigma^-(v=1)$\footnote{Constants constrained by lines with $N\leq 5$.}}     \\
      $ T = 83813.5024(59)  $\qquad   % new Nelson data 2017-05-26 calibrated down 0.06
      $ B = 8.0568(59)     $       \\
      $ D = 0.00076(21)    $\qquad
      $\gamma = -0.1324(91)$ \\
      \bottomrule
    \end{tabular}
  \end{minipage}
\end{table*}

\begin{table}
  \caption{Experimental term values of OH $D\,{}^2\Sigma^-(v=0)$ (cm$^{-1}$, $1\sigma$ uncertainties in parentheses).}
  \label{tab:OH_D00_terms}
  \centering
  \begin{tabular}{ccc}
    \toprule
    $N$ &$T(F_1)$                 & $T(F_2)$                \\
    \midrule
    0   & 81798.350(17)           & --                      \\
    1   & 81828.595(13)           & 81829.121(20)           \\
    2   & 81889.225(13)           & 81890.088(19)           \\
    3   & 81980.148(18)           & 81981.361(28)           \\
    4   & 82101.163(21)           & 82102.528(27)           \\
    5   & 82252.406(26)           & 82254.097(21)           \\
    6   & 82433.505(50)           & 82435.546(41)           \\
    7   & 82644.012(94)           & 82646.507(72)           \\
    8   & 82884.19(21)\phantom{0} & 82886.34(16)\phantom{0} \\
\bottomrule
\end{tabular}
\end{table}

We have used effective Hamiltonians to fit the transitions appearing in our spectra with their parameters listed in Table~\ref{tab:Hamiltonian constants}.
Individually-fitted rotational energies for the perturbed $D(0)$ level of OH are listed in Table~\ref{tab:OH_D00_terms}.
For completeness, molecular parameters are included in Table~\ref{tab:Hamiltonian constants} for OH $X(0)$, $A(0)$ and $D(1)$ states, and OD $X(0)$ and $A(0)$ states that are taken directly or deduced from previous experimental works, e.g., \cite{coxon1980,amiot1981,de_beer1991,stark1994,bernath2009}.
These earlier studies mostly access higher rotational levels than is done here, although with less precision, and consulting these for information on higher-$J$ energy levels will likely give more accurate data than the parameters in Table~\ref{tab:Hamiltonian constants}.
 
An absolute calibration of $D(0)$ and $D(1)$ term origins was made with respect to the $A(0)\leftarrow X(0)$ transitions measured by \citet{stark1994}, who claim an absolute accuracy of \np[cm^{-1}]{0.0005}.

Transition frequencies and $f$-values of individual lines and levels may be of more immediate use than effective Hamiltonians and we have included a complete set of these in tabular form in an online appendix to this paper.

\subsection{Oscillator strengths}

\begin{table}
  \centering
  \begin{minipage}{\linewidth}
    \caption{Band $f$-value ratios and absolute values.\protect\footnote{$1\sigma$ uncertainties in parenthesis in units of the least significant digit.}}
    \label{tab:fvalues}
    \small
    \begin{tabular}{llll}
      \toprule
      Band
      & Source\footnote{Source of $f$-values: RF or DC discharge experiments, or a $N=0$ calculation using the \emph{ab initio} potential-energy curves and transition moments of \protect\citet{van_der_loo2005}.}
      & $\frac{f_{v'v''}}{f_{A(0)\leftarrow X(0)}}$
      & $f_{v'v''}$\footnote{Calibrated against the OH $A(0)\leftarrow X(0)$ band $f$-value measured by \textcite{wang1979} extrapolated to $J''=0$. The same $f$-value is used to calibrate bands of OD but with increased uncertainty.}
      \\
      \midrule
      OH $A(0)\leftarrow X(0)$ &           & 1         & 0.00113(2) \\
      OH $D(0)\leftarrow X(0)$ & RF        & 12.10(87) & 0.0135(10)   \\
      & DC\,I     & 15.2(12)  &            \\ %1504
      & DC\,II    & 13.4(12)  &            \\ %1603
      & calc.     & 11.2      &            \\
      OH $D(1)\leftarrow X(0)$
      & DC\,I / RF\footnote{Calculated as the product of the DC~I $f_{D(1)\leftarrow X(0)}/f_{D(0)\leftarrow X(0)}$ and RF $f_{D(0)\leftarrow X(0)}/f_{A(0)\leftarrow X(0)}$ ratios.}
      & 1.94(20)  & 0.0021(3) \\ %1504
      & calc.     & 2.18      &            \\
      \\ 
      OD $A(0)\leftarrow X(0)$ &           & 1         & 0.0011(1)  \\
      OD $D(0)\leftarrow X(0)$ & RF        & 9.96(73)  & 0.0114(13)   \\
      & calc.     & 9.96      &            \\
      OD $D(1)\leftarrow X(0)$ & RF        & 2.76(25)  & 0.0031(4) \\
      & calc.     & 2.89      &            \\
      \bottomrule
    \end{tabular}
  \end{minipage}
\end{table}

The measured ratios of $D(0)\leftarrow X(0)$ and $D(1)\leftarrow X(0)$ $f$-values with respect to $A(0)\leftarrow X(0)$ are listed in Table~\ref{tab:fvalues}.
The listed uncertainties include an estimate of the fitting uncertainty and an estimated 7\% uncertainty due to possible variation of the OH and OD column densities during the course of the measurements.

As a check on the experimental $f$-value ratios, Table~\ref{tab:fvalues} also lists theoretically-estimated ratios that 
show good agreement given the experimental uncertainty.
We calculate these from the \emph{ab initio} $A\,{}^2\Sigma^+$, $D\,{}^2\Sigma^-$ and $X\,{}^2\Pi$ potential-energy curves of \citet{van_der_loo2005} as well as their $A-X$ and $D-X$ electric-dipole transition moments.
We used standard methods and a custom code similar to \textcite{le_roy2017} which solves the radial Schr\"odinger equation for ground and excited levels providing wavefunctions of internuclear distance, $R$, and integrate the overlap of these wavefunctions weighted by the $R$-dependent $A-X$ and $D-X$ transition moments.
This provides a simple estimate of band strength \cite{luque1998,le_roy2017} but without consideration of the spin-orbit splitting of the doublet states involved.
The same potential-energy curves and electronic transition moment are used for both OH and OD but with the reduced mass adjusted appropriately when solving the radial Schr\"odinger equation.
A more detailed theoretical treatment of $A(v')\leftarrow X(v'')$ absorption $f$-values was made by \textcite{luque1998}, including the deduction of a semi-empirical electronic transition moment.

The experimental relative $f$-values are placed on an absolute scale by reference to the OH\,$A(0)\leftarrow X(0)$ $f$-value measurement of \textcite{wang1979}, who also find a slight decrease of band $f$-value with increasing $J''$ and fit this to the formula $f_{v'v''}=(1.13\pm 0.01)\times 10^{-3} - (1.1\pm0.2)\times 10^{-6}J''(J''+1)$.
This corresponds to a 20\% reduction by $J''=29/2$, which is not at all evident in our relatively low-$J$ measurements.
\textcite{luque1998} calculate an OH\,$A(0)\leftarrow X(0)$ band $f$-value of 0.00104 (for $N'=4$) that is 3\% smaller than the value deduced by \textcite{wang1979} and based on experimental lifetimes \cite{dimpfl1979,german1975b} with about 2\% uncertainty.

To determine whether large $J''$ dependencies affect our measurement of $D(v')-X(0)$ $f$-values, we extend our calculation of OH band strengths to excited rotational transitions by adding centrifugal barriers to the potential-energy curves of the ground and excited states.
This reproduces the $J''$-dependent slope of \textcite{wang1979} for $A(0)\leftarrow X(0)$ within 20\% and finds similar or lesser $J$ dependencies for the other observed bands.
Thus, the experimental band $f$-value ratios listed in Table~\ref{tab:fvalues} are representative of their average value over the range $0.5\lesssim J \lesssim 11.5$ and in accord with a rotationless value within their experimental uncertainty.

We also calculate the OH and OD $A(0)\leftarrow X(0)$ band $f$-values to be in agreement within 1\% for $J=0$, as is also found experimentally by  \textcite{becker1973} (with 10\% uncertainty) and in the calculation of \textcite{luque1998} (as published in the LIFBASE database) when averaged over transitions with $J''=0.5$ to 11.5.
Given this, we scale our OD relative $f$-values measurements to the same reference $A(0)\leftarrow X(0)$ $f$-value as for OH, and conservatively adopt a larger systematic uncertainty of 10\%.

A discrepancy exists between our three independently-measured $D(0)\leftarrow X(0)$/$A(0)\leftarrow X(0)$ $f$-value ratios for OH.
Two of these ratios (DC\,I and DC\,II) were recorded 10 months apart utilising the DC-discharge radical source and both measurements are larger than a value recorded with the RF-discharge source 6 months subsequently, given their estimated uncertainties.
This suggests some unaccounted systematic uncertainty is present or there is an unexpected variability of the column densities maintained by the DC radical source.
We have greater confidence in the RF experiment due to its better thermal stability, and production of a greater column density of radicals and resultant signal-to-noise ratio, allowing for  better spectral resolution.
We then adopt this $f$-value ratio as our best value.
The $f$-value ratio of $D(1)\leftarrow X(0)$ to $D(0)\leftarrow X(0)$ will not be affected by moderate amounts of source variability because these are recorded simultaneously within a single undulator band pass.
The DC-source measurement of this ratio is then combined with RF data to determine an absolute $D(1)\leftarrow X(0)$ $f$-value in Table~\ref{tab:fvalues}.

\begin{table}
  \caption{Experimental and theoretical OH $D\,{}^2\Sigma^-(v'=0)\leftarrow X\,{}^2\Pi(v''=0)$ band $f$-values.}
  \label{tab:OH D(0)-X(0) fvalues}
  \begin{minipage}{\linewidth}
    \centering
    \begin{tabular}{cl}
      \toprule
      \multicolumn{1}{c}{$f_{v'v''}$}                   & Reference                                                                                                                                                                                                                                                                                                                                            \\
      \midrule $~~~~~~~0.0135(10)$ & This work                                                                                                                                                                                                                                                                                                                                            \\
      $0.008-0.013$         & \textcite{nee1984}\footnote{Low-resolution spectrum recorded in the flash photolysis of \ce{H2O}. The smaller value neglects an observed absorbing continuum which may not arise from OH.}                                                                                                                                                           \\
      0.015                 & \textcite{chaffee1977}\footnote{Observed in three sight lines through diffuse interstellar clouds with an uncertainty of about 25\%. This $f$-value was calibrated with respect to the known $A(0)\leftarrow X(0)$ $f$-value at the time \cite{herbig1968} and we have rescaled it to subsequent and more accurate reference data \cite{wang1979}. } \\
      0.012                 & \textcite{van_dishoeck1983b}\footnote{An \emph{ab initio} calculation with a 20\% uncertainty estimated by the authors.}                                                                                                                                                                                                                             \\
      0.013                 & Calculation\footnote{Calculated by us using the potential-energy curves and transition dipole moment of \protect\citet{van_der_loo2005}, as discussed in the text.}                                                                                                                                                                      \\
      \bottomrule
    \end{tabular}
  \end{minipage}
\end{table}
The new OH $D(0)\leftarrow X(0)$ $f$-value is compared with previous determinations in Table~\ref{tab:OH D(0)-X(0) fvalues} that all probe low-$J$ transitions or are rotationless calculations.
These are in very good agreement, suggesting this quantity is now very well established.
Our measured OH $D(1)\leftarrow X(0)$ $f$-value, $0.0021\pm 0.0003$ is somewhat smaller than the calculated value of \citet{van_dishoeck1983b}, 0.0044, and in good agreement with a value we calculate from the \emph{ab initio} potential-energy curves and transition moments of \textcite{van_der_loo2005}, 0.0025.
A list of individual line $f$-values taking into account their rotational line strength factors for all observed transitions is given in the online appendix.

\subsection{$D\,{}^2\Sigma^-(v=0)$ and $D\,{}^2\Sigma^-(v=1)$ linewidths}
\label{sec:linewidths}

\begin{table}
  \caption{Natural linewidths of $D\,{}^2\Sigma^-(v)$ levels.}
  \label{tab:D_linewidths}
  \begin{minipage}{\linewidth}
    \centering
    \begin{tabular}{ccc}
      \toprule
      Species & Level & Linewidth\footnote{In units of cm$^{-1}$\,FWHM with estimated $1\sigma$ uncertainties given in parentheses in units of the least-significant digit.} \\
      \midrule
      OH      & $v=0$ & See Fig.~\ref{fig:D(0) linewidths} \\
      OH      & $v=1$ & 0.25(10) \\
      OD      & $v=0$ & 0.25(05) \\
      OD      & $v=1$ & 0.29(10) \\
      \bottomrule
    \end{tabular}
  \end{minipage}
\end{table}

It was noted by \textcite{van_dishoeck1983b} that some molecules photoexcited into bound levels of the $D\,{}^2\Sigma^-$ state may radiatively decay into the lower-lying unbound excited state $1\,{}^2\Sigma^-$ and then dissociate.
They determined a branching ratio for this process of 20\%, with most molecules safely radiating to the ground state.
They also point out that nonradiative decay of $D\,{}^2\Sigma^-$ levels may dominate this process.
A much shorter lifetime was subsequently predicted for OH $D(0)$ and $D(1)$ by \textcite{van_der_loo2005}.
They calculate the rate of nonradiative decay of these levels by rotational and spin-orbit coupling to co-energetic and repulsive excited states and found significant contributions arising from the $2\,{}^2\Pi$ and $1\,{}^4\Pi$ states, predicting total lifetimes for $D(0)$ and $D(1)$ of \np{1.0e-10} and \np[s]{1.2e-10}, respectively.
These lifetimes imply absorption linewidths of \np[cm^{-1}\,FWHM]{0.05}.

Our measured linewidths for the $D(0)$ and $D(1)$ levels of OH and OD are given in Table~\ref{tab:D_linewidths} and Fig.~\ref{fig:D(0) linewidths}, and essentially give a common value of 0.25\,cm$^{-1}$\,FWHM.
This is a factor of 5 greater than the width of \textcite{van_der_loo2005}, which is perhaps a reasonable agreement given the complexity of their calculation.
An upper limit of 8\,ns was determined in a laser-based lifetime measurement of OH $D(0)$ and is also consistent with our results.
A lower limit on the OH $D(0)$ lifetime (upper limit on its linewidth) was estimated by \textcite{de_beer1991} to be $\np{5e-10}/J(J+1)$\,s, which is incompatible with our measurement for low $N$ values, requiring a linewidth below \np[cm^{-1}\,FWHM]{0.1} for $N$ less than 2.
We can rule out experimental effects and collisional broadening in our experiment due to the positive lack of observed broadening of $A(0)\leftarrow X(0)$.

The observed $N$-dependent increase of OH $D(0)$ linewidths could potentially be the result of its rotational-coupling to the $2\,{}^2\Pi$ state predicted by \textcite{van_der_loo2005}, and a rotational dependence of lesser magnitude cannot be ruled out for the widths of other OH and OD $D$-state levels that we have studied.

\subsection{Local perturbation of OH $D\,{}^2\Sigma^-(v=0)$}
\label{sec: OH D(0) perturbation}

The $D$-state interaction with various repulsive excited-states studied by \textcite{van_der_loo2005} do not explain the observed localised width peak and shifted energy levels of OH $D(0)$  near $N=3$ and 4.
These phenomena  would seem to require a weak interaction with a bound level.
Previous extensive calculations of ${}^2\Pi$, ${}^2\Sigma^+$, and ${}^2\Sigma^-$ states \cite{van_dishoeck1984a} effectively rule out an additional bound level of such a symmetry in the vicinity of $D(0)$, and the repulsive $1{}^4\Pi$ state of \textcite{van_der_loo2005} similarly argues against a state of that symmetry being responsible.
We could find no detailed calculations of further ${}^4\Sigma^+$ and ${}^4\Sigma^-$ states that might support a bound level in the vicinity of $D(0)$ and a perturbing spin-orbit interaction.

\subsection{Interstellar dissociation rate}

The ultraviolet photodissociation rate for OH exposed to a standard Draine interstellar radiation field \cite{draine1978} was calculated using the computed cross section of \textcite{van_dishoeck1984a} that considers a large number of excited states, giving a value of \np[s^{-1}]{3.9e-10} \cite{van_dishoeck2006}.
This rate involves a contribution from the spontaneous radiative decay of bound levels of the $D$ state \cite{van_dishoeck1983b}.
After subtracting that contribution and adding absorption into now 100\%-predissociative $D(0)$ and $D(1)$ levels and using our measured absorption $f$-values this rate is increased to \np[s^{-1}]{4.1e-10}.
This difference is less than the uncertainty introduced by \emph{ab initio} calculation of the direct photodissociation rates of repulsive OH excited states by \textcite{van_dishoeck1984a}.

\section{Summary}

There are four main results from this experimental photoabsorption study targeting the VUV $D\,{}^2\Sigma^-(v'=0)\leftarrow X\,{}^2\Pi(v''=0)$ and $D\,{}^2\Sigma^-(v'=1)\leftarrow X\,{}^2\Pi(v''=0)$ bands of OH and OD, summarised below.
\begin{itemize}
  \item Absolute $f$-values were calibrated with respect to the $A\,{}^2\Sigma^+(v'=0)\leftarrow X\,{}^2\Pi(v''=0)$ band and found to be in agreement with previous laboratory measurements, \emph{ab initio} calculations, and interstellar absorption, with a much improved accuracy.
  The new OH and OD $D\,{}^2\Sigma^-(v'=0)\leftarrow X\,{}^2\Pi(v''=0)$ measurements provide increased confidence when transforming astronomical VUV absorption spectra from interstellar clouds into OH column densities, or calculating the photodissociation rate of OH when exposed to interstellar or stellar VUV radiation.
  The  $D\,{}^2\Sigma^-(v'=1)\leftarrow X\,{}^2\Pi(v''=0)$ $f$-value has not previously been measured.
  \item Predissociation broadening of the observed absorption lines indicate a lifetime of 20\,ps for the $D\,{}^2\Sigma^-(v=0)$ and $D\,{}^2\Sigma^-(v=1)$ levels, and with rotational variation evident in the case of OH $D\,{}^2\Sigma^-(v=0)$.
  This lifetime is 5 times shorter than deduced in a previous comprehensive theoretical study \cite{van_der_loo2005}.
  \item A local perturbation near the $N=4$ level of $D\,{}^2\Sigma^-(v=0)$ in OH is indicated by slight level-energy shifts and increased predissociation widths.
  This is likely due to a weak interaction between $D\,{}^2\Sigma^-$ and another unknown excited state supporting bound levels. 
  \item Line frequencies, $f$-values, and natural linewidths for all observed transitions and level term values are provided in an online appendix.
\end{itemize}

The combination of high-spectral resolution interferometry with broad bandwidth and intense radiation available on the DESIRS photoabsorption branch at SOLEIL has permitted this uniquely-quantitative study of the VUV photoabsorption of OH.
Further refinement of the discharge radical source or development of a window-free version will permit the extension of this work to other molecular radicals and even shorter wavelengths.

\section*{Acknowledgements}
ANH was supported for this work by the postdoctoral fellowship program of PSL Research University.
We are indebted to Denis Joyeux for his help and guidance while running the FTS.
The authors are grateful to Jean-François Gil for the design of the DC and RF cells and for technical assistance during the campaigns on the DESIRS beamline.
We are grateful to the general SOLEIL staff for smooth running of the facility under projects 20140920 and 20160129.

\section*{\refname}
\bibliographystyle{elsarticle-num-names}
\bibliography{article}

\end{document}